\documentstyle[aps,epsf]{revtex}

\begin{document}

\newcommand{\gsim}{\lower.7ex\hbox{$\;\stackrel{\textstyle>}{\sim}\;$}}
\newcommand{\lsim}{\lower.7ex\hbox{$\;\stackrel{\textstyle<}{\sim}\;$}}

\baselineskip 14pt

\title{What are sterile neutrinos good for?\footnote{Talk presented by
the first author at the American Physical Society (APS) Division of
Particles and Fields Conference (DPF'99), hosted by the University of
California, Los Angeles, from January 5-9, 1999.}}
\author{Mitesh Patel and George M. Fuller\footnote{Email addresses:
{\tt mitesh@physics.ucsd.edu}, {\tt gfuller@ucsd.edu}.}}
\address{Department of Physics, University of California, San Diego,
La Jolla, CA 92093-0350}
\maketitle

\begin{abstract}
Taken at face value, current experimental data indicate the existence
of a new particle, the sterile neutrino, which must be a singlet under
the Standard Model gauge group. Although they are not detectable
through traditional means, such particles have interesting
{\it observable} consequences for particle astrophysics and cosmology.
Here we examine these implications and discuss, in particular,
sterile neutrino dark matter and the relationship between
matter-enhanced active-sterile neutrino transformation and the
synthesis of heavy elements in supernovae.
\end{abstract}

\section{Introduction}

In 1930, after inferring the existence of the neutrino from the
continuous electron spectrum of nuclear $\beta$-decay, W.~Pauli
remarked \cite{pauli}, ``I have done a terrible thing. I have
postulated a particle that cannot be detected.'' Twenty-three years
later, F.~Reines and C.~L.~Cowan \cite{reinescowan} reported the first
detection of neutrinos via inverse $\beta$-decay. 

Encouraged by these events, the propensity of history to repeat
itself, and a steady stream of positive experimental data,
contemporary particle physicists and astrophysicists have recently
explored the ramifications of so-called ``sterile''
neutrinos, denoted $\nu_s$. The existence of such
Standard Model-singlet fermions, which couple to the conventional (or
``active'') neutrinos $\nu_e$, $\nu_\mu$, and $\nu_\tau$ solely
through effective mass terms, is implied by the confluence of several
neutrino oscillation experiments:

\begin{itemize}

\item {\bf Atmospheric neutrinos}
~The Super-Kamiokande Collaboration has reported convincing
evidence for the suppression of the flux of $\nu_\mu$ and
$\bar{\nu}_\mu$ produced by cosmic ray collisions with the Earth's
upper atmosphere \cite{superk}. (The measured flux of $\nu_e$ and
$\bar{\nu}_e$ is within expectation.) In particular, there is a
statistically significant zenith angle dependence of the high energy
muon-like events which is consistent with neutrino
oscillations.\footnote{Continued observations and future long
base-line accelerator experiments will help rule out other
explanations for the anomaly. See
Ref. \cite{pakvasa} for a catalog and discussion of these ``non-standard''
solutions.} A two-neutrino vacuum mixing fit yields $\delta
m^2 \approx 10^{-3}-10^{-2}\ {\rm eV}^2$ and $\sin^2(2\theta) \approx
1$, if the neutrino mixing
maximally with $\nu_\mu$ is $\nu_\tau$ \cite{superk}. There are also
matter-enhanced (Mikheyev-Smirnov-Wolfenstein or MSW \cite{msw})
solutions $\delta m^2 \approx \pm 5\times 10^{-3}\ {\rm eV}^2$ and
$\sin^2(2\theta) \approx 1$ if the mixing partner is $\nu_s$
\cite{yasuda}.\footnote{More recent data favors the $\nu_\mu
\rightleftharpoons \nu_\tau$ channel over
$\nu_\mu\rightleftharpoons\nu_s$ \cite{superkrecent}.}

\item {\bf Solar neutrinos}
~An array of solar neutrino experiments \cite{bahcall} has observed an
energy-dependent deficit of $\nu_e$ emitted by nuclear reactions in the
sun. The Kamiokande and Super-Kamiokande experiments observe about
one-half of the expected flux of the highest energy solar neutrinos. The
Homestake chlorine experiment, sensitive to intermediate and higher
energy neutrinos, sees approximately one-third of the expected
flux. Further, the Soviet/Russian-American Gallium Experiment (SAGE)
and Gallex record roughly 
one-half of the expected flux integrated over nearly the entire solar
spectrum. The combined result is a distorted spectrum which is
extremely difficult to reconcile with the standard solar model.
Global two-neutrino fits to these observations
yield \cite{bahcall} the large angle
($\delta m^2 \approx 10^{-5}\ {\rm eV}^2$, $\sin^2(2\theta)
\approx 1$) and small angle
($\delta m^2 \approx 10^{-5}\ {\rm eV}^2$, $\sin^2(2\theta) \approx
5\times 10^{-3}$)
MSW solutions and the ``just-so''
vacuum oscillation solution
($\delta m^2 \sim 10^{-10}\ {\rm eV}^2$, $\sin^2(2\theta) \approx 1$).
The neutrino mixing with $\nu_e$ may be
$\nu_\mu$, $\nu_\tau$, or $\nu_s$, depending on the
solution.\footnote{Of course, requiring compatibility with other
experiments and astrophysical constraints (see below) restricts the
allowed oscillation channels.}

\item {\bf Accelerator neutrinos}
~The Los Alamos Liquid Scintillator Neutrino Detector (LSND)
experiment has recorded an excess of $\nu_e$ and $\bar{\nu}_e$ events
in accelerator-produced beams of $\nu_\mu$ and $\bar{\nu}_\mu$
respectively \cite{athana}. There are several allowed regions in the
oscillation parameter space, all of which fall in the ranges
$0.2\ {\rm eV}^2 \lsim \delta m^2 \lsim 8\ {\rm eV}^2$ and
$10^{-3} \lsim \sin^2(2\theta) \lsim 10^{-1}$,
assuming two-neutrino mixing. The agreement \cite{athana}
between the regions
for the neutrino ($\nu_\mu\rightleftharpoons\nu_e$) and antineutrino
($\bar{\nu}_\mu\rightleftharpoons\bar{\nu}_e$) channels reinforces the
oscillation interpretation. The Karlsruhe Rutherford Medium Energy
Neutrino (KARMEN) experiment has searched for excess events in the
same channels \cite{karmen}, and despite a null result, a joint
analysis of the KARMEN and LSND data preserves some of the LSND
solution space \cite{eitel}. 

\end{itemize}

The mutual incompatibility of these disparate sets of
results is {\it prima facie} evidence for the existence of a {\it
light} sterile neutrino $\nu_s$, since the number of light weakly
interacting neutrino species is known to be three (namely, $\nu_e$,
$\nu_\mu$, and $\nu_\tau$) \cite{pdg}, an effectively three-neutrino
mass matrix
yields at most two independent $\delta m^2$'s, and the active
neutrinos are known to be very light compared to their charged
leptonic counterparts.\footnote{If the additional neutrino is not a
Standard Model (SM)
singlet, then it must have {\it very weak} interactions with the SM
particles in order to meet these constraints. See
Ref. \cite{kolbmohatep} for an astrophysical limit on
such interactions.}
Global fits of the data to three-neutrino mass matrices
have borne out this conclusion, and several authors have indicated how
to accommodate all of the data in a four-neutrino mixing matrix
\cite{bgg}.

Assuming that the current data is explained by oscillations
and that future neutrino experiments confirm the reality of an
additional light
neutrino species, a number of workers have begun constructing
theoretical models which yield the naturally small Dirac {\it and}
Majorana neutrino masses required for appreciable
active-sterile neutrino mixing. Some of these models
are generalizations of the traditional see-saw mechanism \cite{seesaw}
for generating light active neutrinos in the presence of very heavy
sterile neutrinos \cite{mohapatra}. Others involve restricted or
extended couplings in the context of the Standard Model (SM)
\cite{mclaughlinng}. Many methods rely on new symmetries to ensure
that both active and sterile neutrino masses are small and
comparable. All of them can be classified as simple extensions of the
SM gauge group and matter content (including Grand Unified
Theories or GUTs), supersymmetric models, or superstring-inspired
scenarios.

The phenomenological consequences and uses of sterile neutrinos
are equally interesting. In particular, resonant
transitions among active and sterile neutrinos can alter significantly
the dynamics of early universe cosmology and various astrophysical
venues. Indeed, significant consequences are almost guaranteed in
phenomena such as Big Bang nucleosynthesis (BBN) and core-collapse
supernovae, whose outcome is determined or dominated by neutrino
physics. Transitions to and from sterile neutrinos can distort
severely the active neutrinos' energy spectra, resulting, for
example, in nucleosynthetic abundances markedly different from the
commonly accepted values.

These effects are not invariably unfavorable. Indeed, sterile
neutrinos have been variously invoked to explain the origin of pulsar
kicks \cite{kusenko}, provide a new dark matter candidate
\cite{shifuller}, account for the diffuse
ionization in the Milky Way galaxy \cite{diffuse},
resolve the ``crisis'' in BBN \cite{kev}, and
help enable the synthesis of heavy elements in Type II supernovae
\cite{gail,mitesh}. In
these proceedings, we describe the interesting phenomenological
implications of two of these scenarios. We recapitulate in Sec. II the
physics of sterile neutrino dark matter, concentrating on the
production of the cold, non-thermal variety. In Sec. III, we summarize
our recent work on and the status of matter-enhanced active-sterile
neutrino transformation solutions to heavy-element
nucleosynthesis. We give conclusions in Sec. IV.

\section{Sterile neutrino dark matter}

The number of suitable candidates for the dark matter of the universe
is staggering! In Eq. (\ref{omegatotal}) we have listed some of the
possible constituents --- without regard to their potential mutual
exclusivity --- of the total mass-energy of the universe (measured
as a fraction $\Omega$ of the Friedmann-Robertson-Walker (FRW) closure
energy density):\footnote{A number of these dark matter hopefuls have
been culled from Ref. \cite{kolbturner} and preprint archive listings
\cite{archive}. See Ref. \cite{kolbturner} for a lucid discussion of
the dark matter problem and structure formation.}

\begin{eqnarray}
\Omega_{\rm TOTAL} &=&
\Omega_{\rm baryon} + \Omega_{\rm \Lambda} + \Omega_{\rm axion} +
\Omega_{\rm \nu_{e,\mu,\tau}} + \Omega_{\rm Q-ball}
\nonumber\\
&&
+~ \Omega_{\rm LSP} + \Omega_{\rm WIMP} + \Omega_{\rm WIMPZILLA} +
\Omega_{\rm SUSY} + \Omega_{\rm CMBR}
\nonumber\\
&&
+~ \Omega_{\rm monopole} + \Omega_{\rm starlight} + \Omega_{\rm cosmic\
string} + \Omega_{\rm primordial\ black\ hole}
\nonumber\\
&&
+~ \Omega_{\rm \tilde{\chi}^0} + \Omega_{\rm majoron} + \Omega_{\rm
moron} + \Omega_{\rm \tilde{\gamma}} + \Omega_{\rm Newtorite} +
\Omega_{\rm quark\ nugget}
\nonumber\\
&&
+~ \Omega_{\rm \tilde{\nu}} + \Omega_{\rm pyrgon} + \Omega_{\rm \tilde{g}} +
\Omega_{\rm gravitational\ wave} + \Omega_{\rm maximon}
\nonumber\\
&&
+~ \Omega_{\rm \tilde{h}} + \Omega_{\rm familon} + \Omega_{\rm tetron} +
\Omega_{\rm penton} + \Omega_{\rm hexon} + \Omega_{\rm crypton}
\nonumber\\
&&
+~ \bf \Omega_{\nu_s}
\label{omegatotal}
\end{eqnarray}

The astute reader will have noticed the addition of sterile neutrino
dark matter to this list. If one light sterile species
$\nu_s$ is required by the neutrino oscillation experiments, then
the multi-generational structure of the SM argues strongly in favor of
two additional steriles $\nu^\prime_s$ and $\nu^{\prime\prime}_s$. While
more massive than their lighter sibling, these additional neutrinos
could also have relatively small masses which are easily compatible
with the data.

If there is a $\nu^\prime_s$ in the 200 eV to 10 keV mass range and 
a primordial lepton number $L \approx 10^{-3}-10^{-1}$ in any
of the active neutrino flavors, Shi \& Fuller \cite{shifuller}
have shown that this
asymmetry can drive resonant production of sterile neutrino dark
matter in the early universe. Furthermore, since the MSW mechanism
in this environment favors efficient conversion from active to sterile
neutrinos only for the lowest energy neutrinos and the process itself
destroys lepton number, the resulting $\nu^\prime_s$ spectrum is
non-thermal and essentially ``cold.''

Unlike conventional neutrino dark matter (Hot Dark Matter or HDM),
this Cold Dark Matter (CDM) candidate has a relatively short
free-streaming length at the epoch of structure formation, so density
fluctuations may grow unimpeded on galactic scales. For example, a
$\approx 500$ eV sterile neutrino can contribute a fraction
$\Omega_{\nu_s} \approx 0.4$ to the critical density and stream freely
over only $\approx 0.4$ Mpc, which corresponds to a structure cut-off
$\approx 10^{10} M_\odot$, about the size of dwarf galaxies.
For further details, please consult Ref. \cite{shifuller}. 

\section{Active-sterile neutrino transformation and heavy-element
nucleosynthesis}

What is the origin of the heavy elements? The light elements (H, He,
D, and Li) are produced (for the most part) in the early universe and
the intermediate mass elements (up to Fe) during the ``main sequence''
evolution of stars,\footnote{See Ref. \cite{clayton} for a detailed
exposition on stellar structure and dynamics.} but the source of at
least half of the nuclides heavier than iron is unknown.

It is not difficult to understand why ordinary stellar evolution
cannot produce nuclei more massive than iron. The thermonuclear
processes which take place in stellar interiors work gradually from
hydrogen toward iron by fusing together ever heavier nuclei. Since
these thermal reactions can proceed only if the products are more
tightly bound than the reactants, the path of stellar nucleosynthesis
ends at iron, which is the most tightly bound nucleus.

It is clear, then, that elements heavier than iron must form via some
other process(es) in some other environment(s). One pathway from iron
to heavy nuclides such as plutonium, iodine, tin, lead, and gold is
the {\bf r}-process, so named because it proceeds via the {\bf r}apid
capture of
neutrons on iron-sized ``seed'' nuclei. In the r-process, seed nuclei
capture neutrons so rapidly that they can leap past iron on the curve
of binding energy versus atomic mass. After the r-process is complete,
the now extremely neutron-rich nuclei $\beta$-decay to more stable
elements on the periodic table, yielding some mass distribution of
nuclides.\footnote{Preferably a distribution which matches the
observed abundances!} In all but extreme astrophysical environments,
a key necessary condition for a successful
r-process is a preponderance of neutrons over protons, as the heavy
elements are quite neutron-rich.\footnote{Neutron-richness of the
heavy nuclides can be understood roughly as a compromise among Pauli's
exclusion principle, long-range Coulomb forces between protons, and
short-range nuclear (strong) forces between nucleons.} Once seed nuclei
are available for neutron capture to proceed, this requirement becomes
having a large number of free neutrons for each seed nucleus.

Although the physics of the r-process is relatively well-known, the
astrophysical environment in which it takes place has not been
determined conclusively. The major contenders include Type II (or
core-collapse) supernovae and binary neutron star mergers. 
The former is currently the most promising r-process site, and we
now restrict ourselves to a discussion of supernova physics and how
non-standard particle physics like neutrino oscillations can help
enable the r-process.

A Type II supernova results from the catastrophic gravitational
collapse of a massive star \cite{clayton,st}.
Specifically, a star with mass $M \gsim
8 M_\odot$ burns successively heavier fuels throughout its life,
eventually accumulating an inner iron core. As indicated
above, thermal fusion processes cannot extract energy from iron to
provide pressure support against gravity, so
the core eventually collapses under its own gravity. Abetted by the
photodissociation of the iron nuclei, the core rapidly neutronizes via
electron capture on 
protons. When the central core density reaches the saturation density
of nuclear matter, the nucleons touch, and the core ``bounces,''
yielding an outgoing shockwave. Unfortunately, the shock wave
eventually dies out, as it loses all of its energy dissociating the
mantle of the star. Meanwhile, however, thermally produced neutrinos,
trapped in the dense core, diffuse out and
revive the shock (the supernova explodes!).
The remaining neutrinos continue to leak out on a time scale of some
10-20 seconds, driving significant mass loss from the newly formed
proto-neutron star.

It is in this neutrino-driven ``wind'' \cite{qian} that the r-process
is believed to occur \cite{woosley}. The hot proto-neutron star emits
neutrinos of all flavors, with average energies 
$\langle E_{\nu_\delta} \rangle
\approx
\langle E_{\bar{\nu}_\delta} \rangle
\gsim 
\langle E_{\bar{\nu}_e} \rangle
\gsim
\langle E_{\nu_e} \rangle
$, where $\delta = \mu,\tau$ refers to the muon and tau
neutrinos, and comparable luminosities.
They stream nearly freely through a plasma of electrons,
positrons, neutrons, and protons. By exchanging energy with the
cooler plasma, the neutrinos drive the r-process ingredients out of
the gravitational potential well of the neutron star. As a fluid
element of this neutrino-heated ejecta travels away from the surface
of the star, its temperature gradually drops, its velocity increases,
and the following sequence ensues:

\begin{itemize}

\item The reactions

\begin{eqnarray}
\nu_e + n &\rightleftharpoons& p + e^-
\label{nkillrate}
\\
\bar{\nu}_e + p &\rightleftharpoons& n + e^+
\label{pkillrate}
\end{eqnarray}

\noindent
set the ratio $n/p$ of the number densities of neutrons and
protons.\footnote{The reactions $n \rightleftharpoons p + e^- +
\bar{\nu}_e$ contribute only negligibly to $n/p$ for the relevant
neutrino energies.}
This number is usually written in terms of the electron fraction
$Y_e$, the net number of electrons per baryon. Charge neutrality of
the plasma gives $Y_e = 1/(1+n/p)$. Thus a higher (lower) $n/p$
corresponds to a lower (higher) $Y_e$. The reactions in Eqs. (2-3)
freeze out at a temperature $T \approx 0.8\ {\rm MeV}$, when their
rates fall below the local expansion rate.

\item When the plasma cools to $T \approx 0.75\ {\rm MeV}$,
$\alpha$-particles ($^4{\rm He}$ nuclei) start to form from the
ambient neutrons and protons. 

\item At lower plasma temperatures, some of the $\alpha$-particles
combine via three-body interactions into seed nuclei with mass number
$A = 50-100$.

\item Finally, at sufficiently low temperatures ($T \approx 0.25$ MeV),
the r-process occurs.

\end{itemize}

Unfortunately, detailed numerical simulations \cite{alphaeffect} have
found that the r-process is {\it precluded} in neutrino-heated ejecta
by a phenomenon termed the ``$\alpha$-effect.'' The $\alpha$-effect
occurs when nucleons are taken up into $\alpha$-particles, leaving
nearly inert $^4{\rm He}$ nuclei and excess neutrons (the plasma
is already neutron-rich). Since the $\nu_e$ flux is still sizable in
the region beyond freeze-out, the forward reaction in Eq. (2)
proceeds to convert the leftover neutrons into protons. These protons
and remaining neutrons quickly form additional $\alpha$-particles
before the forward reaction in Eq. (3) can reconvert the newly-formed
protons. Ultimately, nearly all of the free neutrons end up in
$^4{\rm He}$ nuclei, the electron fraction is driven to 0.5, and the
final neutron-to-seed ratio is too small to give even an anemic
r-process!

We have shown in Fig. 1 the radial evolution of the electron fraction
in the
hot, high-entropy ``bubble'' above the proto-neutron star. We have
used a simple model \cite{alphaeffect} for the wind and taken
representative values of the neutron star radius ($R = 10\ {\rm
km}$), expansion time scale ($\tau = 0.3\ {\rm s}$), entropy per
baryon ($s/k_B = 100$), and neutrino spectral parameters
\cite{mitesh,gail}. The $\alpha$-effect begins when the temperature in
the plasma drops to $\approx 0.75\ {\rm MeV}$. Comparison of the
behavior with and without the $\alpha$-effect illustrates its
negative impact.

\begin{figure}[]     
\epsfverbosetrue
\centerline{\epsfbox{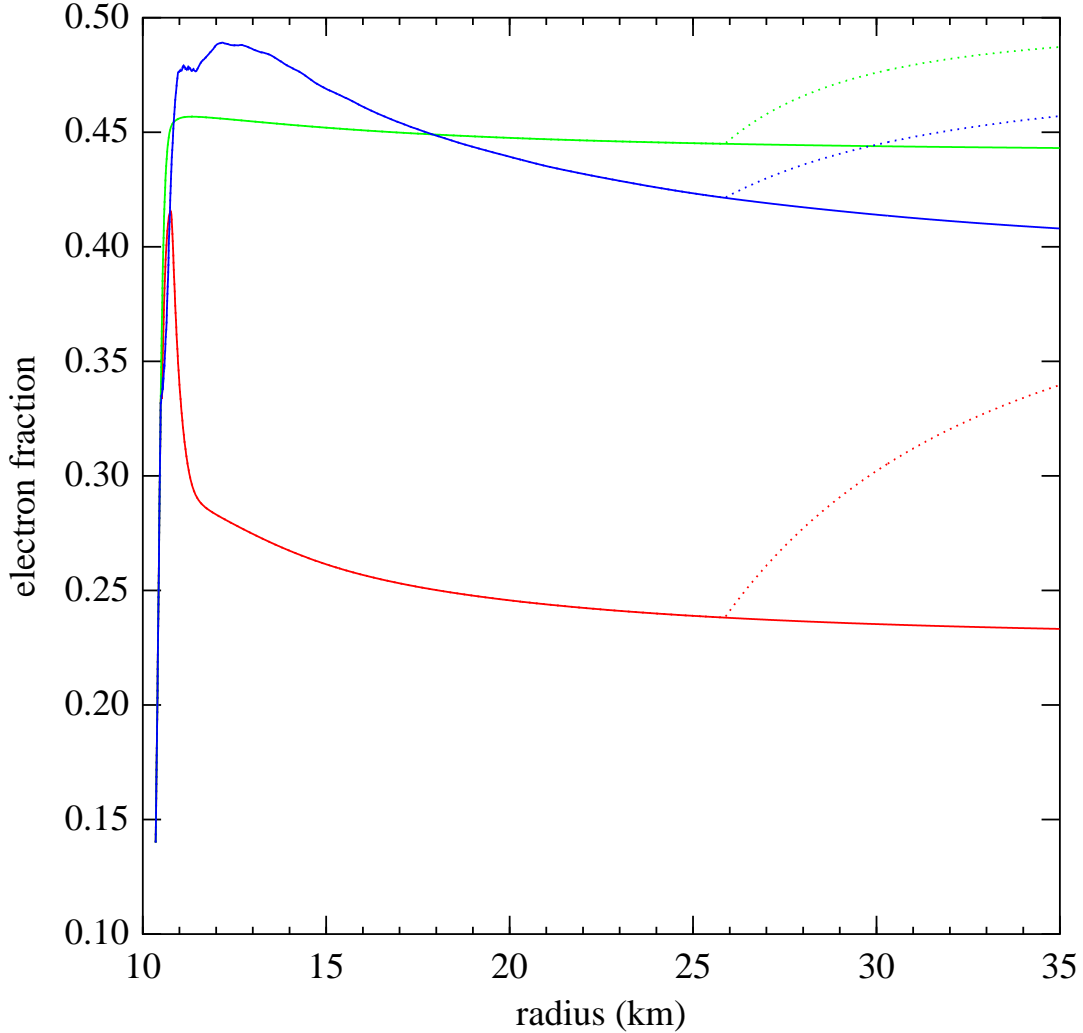}}
\vskip 0.5 cm
\caption[]{
\label{yeprofiles}
\small Evolution of the electron fraction $Y_e$ with radius in the
neutrino-driven wind above
a proto-neutron star for no neutrino oscillations (green),
$\nu_e\rightleftharpoons\nu_s$ and
$\bar{\nu}_e\rightleftharpoons\bar{\nu}_s$
oscillations ($\delta m^2 = 10\ {\rm
eV}^2$, $\sin^2(2\theta) = 0.01$) without
neutrino background effects (red), and such oscillations
with neutrino background effects (blue). The solid curves
indicate the behavior without the $\alpha$-effect and the dotted
curves the behaviour with the $\alpha$-effect. See the text for
the relevant physical parameters.}
\end{figure}

Various calculations \cite{alphaeffect,gail,mitesh} have shown that
this problem persists for all reasonable variations in wind outflow
and neutrino spectral parameters. Since there is mounting evidence
\cite{alphaeffect,gailgail,rprocevidence1,rprocevidence2} that the
r-process must occur in neutrino-heated ejecta in Type II supernovae, a
number of authors \cite{gail,mitesh,cfq} have attempted to circumvent
the $\alpha$-effect in order to enable heavy-element nucleosynthesis.

The high neutron-to-seed ratio requirement for a successful r-process
is met if $Y_e$ is sufficiently small ($Y_e < 1/2$
corresponds to neutron-rich ejecta); the expansion rate of the ejecta
is sufficiently high that the three-body reactions making seed nuclei
are inefficient; and/or the entropy of the plasma is high enough that
nucleons prefer to remain free rather than bound in nuclei
\cite{mbandhwq}. Increasing
the expansion rate will not redress the $\alpha$-effect, since
$\alpha$-particles form via relatively fast two-body
processes. Raising the entropy in the ejecta substitutes for the
$\alpha$-effect another process which decimates the neutron-to-seed
ratio: neutrino neutral current spallation of $^4{\rm He}$ nuclei
\cite{meyercatastrophe}. Since the severity of the effect depends on
the $\nu_e$ flux in the region of $\alpha$-particle production,
lowering $Y_e$ without changing the $\nu_e$ flux also cannot help
enable the r-process.

\begin{figure}[]     
\epsfverbosetrue
\centerline{\epsfxsize 7.0 truein \epsfbox{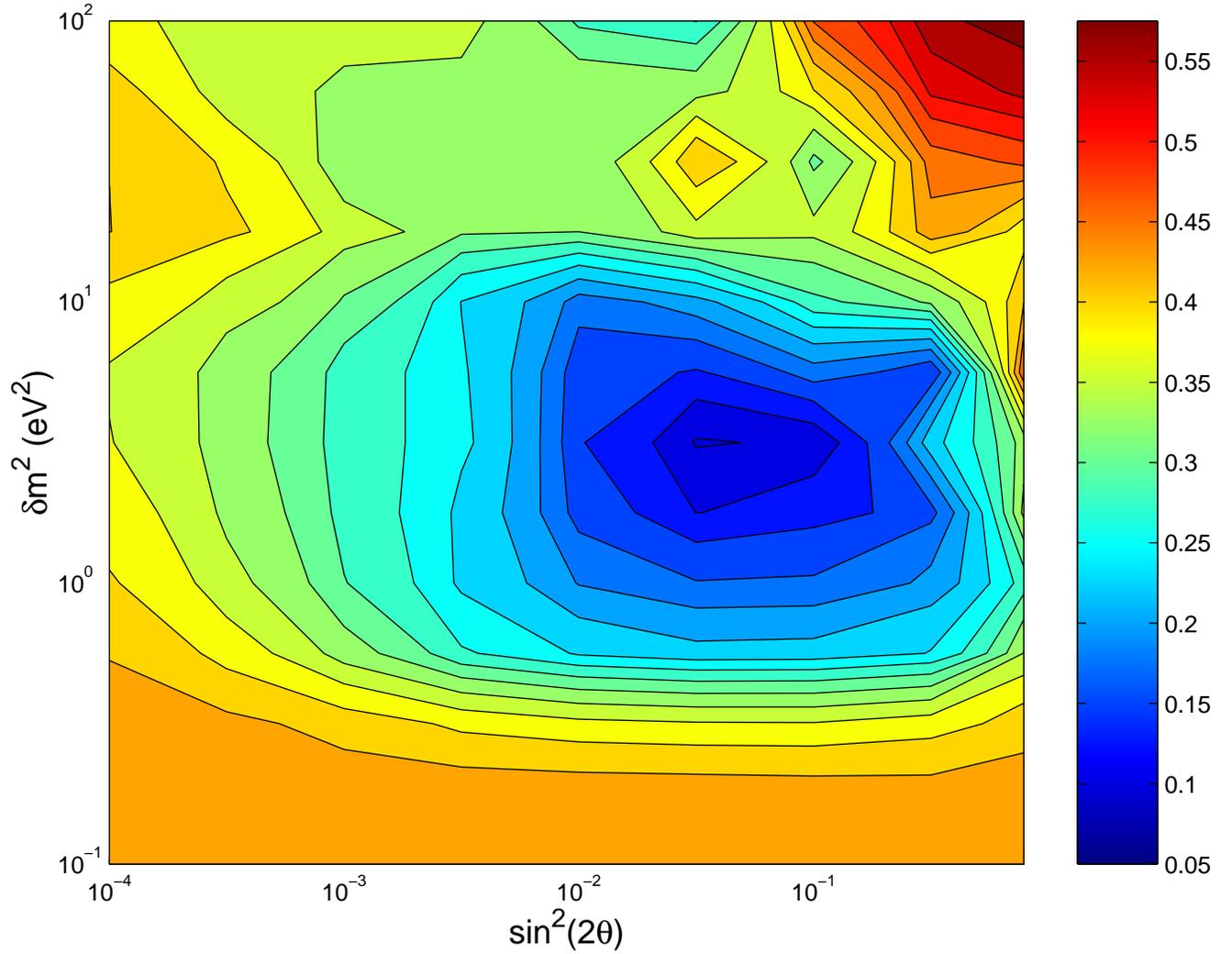}}
\vskip 0.5 cm
\caption[]{
\label{finalyes}
\small Final electron fraction $Y_e$, excluding neutrino background
effects, for various $\nu_e\rightleftharpoons\nu_s$ and
$\bar{\nu}_e\rightleftharpoons\bar{\nu}_s$ oscillation parameters
$\delta m^2$ and $\sin^2(2\theta)$. The color bar at the right
indicates the value of $Y_e$. The physical conditions are the same as
in Fig. 1.}
\end{figure}

Non-standard neutrino physics such as neutrino oscillations is an
elegant solution to this problem, for resonant transformation between
$\nu_e$ and some other species can directly influence the $\nu_e$
flux. Given the hierarchy of the energies of the active neutrinos,
matter-enhanced transformation between $\nu_e$ and $\nu_\mu$ or
$\nu_\tau$ will actually enhance the $\alpha$-effect. Transformations
among the active antineutrinos will affect only
$\bar{\nu}_e$. As discussed in the Introduction, however, various
neutrino experiments imply the existence of a sterile neutrino which
mixes appreciably with the active neutrinos. As long as the
neutron star emits a negligible number of sterile neutrinos,
effective active-sterile mixing in the form of
$\nu_e\rightleftharpoons\nu_s$ and
$\bar{\nu}_e\rightleftharpoons\bar{\nu}_s$ has the potential to forestall
the $\alpha$-effect by removing the offending $\nu_e$'s
\cite{gail,mitesh,cfq}.

The authors of Ref. \cite{gail} and we \cite{mitesh} have recently
investigated this mixing scheme in neutrino-heated ejecta. Shown in
Fig. 1 is the evolution of $Y_e$ with radius for a representative set
of mixing parameters. In Ref. \cite{gail}, the authors included all
contributions to the effective mass \cite{ng} of electron neutrinos
except neutrino forward scattering on the ``background'' of other
neutrinos emitted by the neutron star \cite{qianback}.
In Ref. \cite{mitesh}, we have also
included this many-body effect. The behavior of $Y_e$ for both
analyses is indicated in the figure, both with and without the
$\alpha$-effect. In Fig. 2, we have shown final electron fraction
attained (at $r = 35\ {\rm km}$ in Fig. 1) without neutrino background
effects.

In the absence of the neutrino background, active-sterile mixing is a
viable solution to the r-process problem. For a relatively large
range of neutrino mixing parameters, the final electron fraction is
less than 0.3, yielding a highly neutron-rich r-process
environment. Calculations with various choices of the outflow and
neutrino spectral parameters have confirmed that this is also a {\it
robust} solution \cite{gail,mitesh,cfq}: a high neutron-to-seed ratio
obtains for a wide range of supernova and mixing parameters. An
additional benefit is that the $\delta m^2$-$\sin^2(2\theta)$ region
favored for enabling the r-process coincides with one of the
four-neutrino mixing schemes explaining the current experiments
\cite{gail,bgg}.

The presence of the neutrino background unfortunately mitigates
this remedy, but the background gradually dies away as
neutrinos diffuse out of the neutron star. At sufficiently late times
after the supernova bounce and explosion,
the probability of neutrino-neutrino
forward scattering is sufficiently small that the scenario of
Ref. \cite{gail} prevails. At these times, however, the neutrino
fluxes have fallen sufficiently that there may not be a need for
neutrino transformation to relieve the $\alpha$-effect. If the total
neutron-rich mass ejected at late times is sufficient to account for
all of the r-process material in the galaxy, then there is no need to
invoke non-standard neutrino physics. Settling this issue requires a
self-consistent calculation of the wind coupled with neutrino
oscillations and including neutrino background effects. On the other
hand, if future experiments conclusively establish the existence of a
light sterile neutrino in the mass range relevant for Type II
supernovae, active-sterile mixing may have profound consequences for
the synthesis of the heavy elements.

\section{Conclusion}

The present solar, atmospheric, and accelerator neutrino experiments
suggest the existence of a light sterile neutrino. While this is a
radical departure from the folklore that sterile neutrinos (if they
exist) are very heavy, future experiments \cite{bgg} may confirm
their reality and force theorists to modify or discard cherished
models of neutrino mass. As we have indicated in this paper, the
mixing of sterile and active neutrinos has potentially far-reaching
consequences for cosmology and astrophysics. They may account for much
of the dark matter of the universe. They may even be the reason why we
have gold rings, tin cans, atomic bombs, and lead shielding!

MP is supported in part by a NASA GSRP fellowship.
This work was partially supported by NSF grant PHY98-00980.


\end{document}